\documentclass[11pt]{article}

\usepackage[preprint]{acl}

\usepackage{times}
\usepackage{latexsym}

\usepackage[T1]{fontenc}

\usepackage[utf8]{inputenc}

\usepackage{microtype}

\usepackage{inconsolata}
\usepackage{booktabs}
\usepackage{multirow}
\usepackage{makecell}

\usepackage{graphicx}

\usepackage{amsmath}
\usepackage{amssymb}
\usepackage{cleveref}

\usepackage{nameref}

\title{Chatterbox-Flash: Prior-Calibrated Block Diffusion for Streaming Zero-Shot TTS}

\author{Deokjin Seo\thanks{\,\,Equal contribution.} \\
  Resemble AI \\
  \texttt{deokjin.seo@resemble.ai} \\
  \texttt{ejrwls012@gmail.com}
  \And
  Gangin Park\footnotemark[1] \\
  Seoul National University \\
  \texttt{ssonpull519@gmail.com}
  \And
  Kihyun Nam \\
  KAIST \\
  \texttt{nkh.mmai@kaist.ac.kr}}

\begin{document}
\maketitle
\begin{abstract}
We present Chatterbox-Flash, a zero-shot text-to-speech model obtained by fine-tuning a pretrained autoregressive TTS decoder into a block-diffusion decoder, enabling parallel token generation within each block while retaining block-by-block streaming.
We find that naively transferring mainstream block-diffusion decoding to discrete speech tokens degrades quality, as a long-tail token distribution biases parallel position selection toward a few high-frequency tokens.
To mitigate this without architectural modification, we introduce two inference-time techniques: prior-calibrated scoring, which subtracts the block-level marginal token distribution, and an early-decoding schedule, which adaptively terminates iteration based on calibrated confidence.
On standard zero-shot TTS benchmarks, Chatterbox-Flash attains high-fidelity synthesis comparable to strong autoregressive and non-autoregressive baselines, while supporting streaming inference with time-to-first-packet on par with streaming AR systems and substantially lower real-time factor.
Code is available at \url{https://anonymous.4open.science/r/chatterbox-flash-B458}.
\end{abstract}

\section{Introduction}



Zero-shot text-to-speech (TTS), capable of synthesizing speech in unseen speakers' voices from a short reference audio clip, has emerged as a central capability of modern speech synthesis, driven by large-scale multi-speaker training~\cite{wang2023neuralcodeclanguagemodels, le2023voicebox, borsos2023audiolm, anastassiou2406seed}.

Recent zero-shot TTS models can be broadly categorized along two axes.
Along the axis of generation order, autoregressive (AR) models generate tokens sequentially, conditioning each step on previous outputs~\cite{wang2023neuralcodeclanguagemodels, du2024cosyvoice2, du2025cosyvoice, zhou2026indextts2}, while non-autoregressive (NAR) models produce all positions in parallel~\cite{eskimez2024e2, chen2025f5, wang2025maskgct}.
Along the axis of modeling space, they operate either on discrete tokens derived from neural audio codecs~\cite{defossez2022high, zeghidour2021soundstream} or on continuous latent representations~\cite{shen2024naturalspeech}.
Among these combinations, AR language models over discrete audio codecs have proven particularly effective, delivering strong speech quality while remaining amenable to streaming inference~\cite{yang2026measuring}.

Despite these advantages, AR decoding incurs an inherent latency since tokens are produced sequentially and wall-clock time grows linearly with output length---a bottleneck that cannot be removed by engineering alone, motivating model-native parallel decoding.
Among such approaches, diffusion language models (DLMs)~\cite{austin2021d3pm, lou2024sedd, sahoo2024mdlm} generate multiple tokens per step and have recently achieved throughput far above AR models with minimal quality loss, with the LLaDA family~\cite{nie2026large, bie2025llada2, bie2026llada2} and other extensions~\cite{li2026lavida, ye2025dream} scaling the paradigm to discrete large language models (dLLMs).
Block Diffusion~\cite{blockdiffusion} adds block-causal training and inference compatible with streaming, later made practical at scale by Fast-dLLM~\cite{fastdllm} and Fast-dLLM~v2~\cite{wu2025fastdllmv2efficientblockdiffusion}.

However, DLMs remain underexplored for speech, with existing efforts limited in scale~\cite{ku2026discrete}, lacking native streaming~\cite{zhu2026omnivoice}, or, in a similar block-wise paradigm, pairing masked diffusion with a separately-parameterized AR module that only passes a compact hidden state rather than tokens~\cite{song2025distar}.
Moreover, we find that DLM decoding techniques do not transfer directly to discrete speech: codec sequences are heavily skewed toward a few dominant tokens---notably silence---that carry little context-dependent information~\cite{compressed2fine2025, sicherman2023analysing}, and block-by-block decoding restricts position selection to a small local window~\cite{shu2026deferred} where ranking is fragile and not directly supervised~\cite{whereunmask2026}.

In this work, we present Chatterbox-Flash, a zero-shot TTS model obtained by fine-tuning a pretrained autoregressive decoder into a discrete block-diffusion decoder, retaining the original architecture and replacing only the training objective with masked denoising.
To address the problem with degraded quality, we introduce two inference-time techniques---\emph{prior-calibrated scoring} and an \emph{early-decoding schedule}---and integrate them with a block-causal streaming pipeline that delivers streaming inference at substantially lower latency than AR baselines.

Our contributions are as follows:
\begin{itemize}
    \item \textbf{Streaming Block-diffusion TTS} 
    To our knowledge, the first zero-shot TTS model to perform block diffusion decoding in the sense of~\cite{blockdiffusion} — i.e., a language model that directly generates discrete codec tokens under block-causal attention over the full prefix — with native block-by-block streaming.
    \item \textbf{Prior-Calibrated Scoring} An inference-time correction that suppresses the long-tail token bias in parallel position selection, requiring no architectural change or additional forward pass.
    \item \textbf{Early-Decoding Schedule} An adaptive termination rule that lowers the average number of denoising steps below the maximum budget based on calibrated confidence.
    \item \textbf{Empirical Validation} Chatterbox-Flash matches strong AR and NAR baselines in quality while being the only system in our comparison with native streaming support.
\end{itemize}

\section{Method}
\label{sec:method}

\subsection{Modeling}
\label{sec:modeling}

\paragraph{Architecture}
Our system extends Chatterbox-TTS~\cite{chatterboxtts2025}\footnote{\url{https://github.com/resemble-ai/chatterbox}}, an open-source two-stage zero-shot TTS pipeline.
Stage~1 is a Llama-style Transformer decoder (T3) that performs next-token prediction over a discrete speech token sequence $\mathbf{y} = (y_1, \ldots, y_T)$ extracted at 25\,Hz by a neural audio codec.
The decoder is conditioned on
\begin{equation}
\mathbf{c} = [\mathbf{e}_s, \mathbf{x}_{\text{text}}, \mathbf{x}_{\text{speech}}],
\label{eq:cond}
\end{equation}
which combines a global speaker embedding $\mathbf{e}_s$ obtained from a GE2E-trained voice encoder~\cite{wan2020generalizedendtoendlossspeaker}, the input text token sequence $\mathbf{x}_{\text{text}}$, and the prompt speech tokens $\mathbf{x}_{\text{speech}}$ extracted from the reference audio.
The speech distribution is factorized autoregressively as
\begin{equation}
    p(\mathbf{y} \mid \mathbf{c}) = \prod_{t=1}^{T} p(y_t \mid y_{<t}, \mathbf{c}).
    \label{eq:ar}
\end{equation}
Stage~2 is a flow-matching vocoder that converts the generated tokens to waveforms with chunk-wise streaming; it follows the CosyVoice2~\cite{du2024cosyvoice2} architecture, distilled into a few-step flow-matching sampler with MeanFlow~\cite{geng2025meanflowsonestepgenerative}. We reuse this vocoder unchanged and focus our contribution on the Stage-1 decoder, so quality differences across decoding configurations are attributable to

\paragraph{Block Diffusion}
Following~\cite{blockdiffusion, wu2025fastdllmv2efficientblockdiffusion}, we apply discrete denoising diffusion block-by-block on the T3 decoder instead of over the full sequence.
A length-$T$ sequence $x$ is partitioned into $B = \lceil T/D \rceil$ non-overlapping blocks $x^{(1)}, \ldots, x^{(B)}$ of size $D$, giving
\begin{equation}
p(x) = \prod_{b=1}^{B} p\!\left(x^{(b)} \mid x^{(<b)}\right).
\label{eq:block}
\end{equation}
Each factor is modeled by a parallel masked predictor: positions in $x^{(b)}$ are randomly replaced with $[\textsc{m}]$ to form $x_t^{(b)}$, and the predictor recovers the masked tokens in parallel from $x^{(<b)}$ and $x_t^{(b)}$. The left-to-right inter-block factorization naturally supports block-wise streaming generation.

\subsection{Training}
\label{sec:training}

\paragraph{Packed Input and Hybrid Attention}
The packed input $[\mathbf{c}, x_t]$ uses a hybrid attention scheme over $B$ speech blocks (Figure~\ref{fig:attn_mask}): causal over the conditioning $\mathbf{c}$, bidirectional within each speech block, and causal across blocks, so that $x_t^{(b)}$ attends only to $\mathbf{c}$ and $x^{(<b)}$.
Unlike Fast-dLLM~v2~\cite{wu2025fastdllmv2efficientblockdiffusion}, which is block-causal throughout, we keep $\mathbf{c}$ causal to preserve the pretrained backbone's embedding space and apply block-diffusion attention only to the speech part, maintaining the monotonic text-to-speech alignment.
We implement this with custom attention kernels (Appendix~\ref{app:attn_kernel}).

\paragraph{Complementary Masking}
At each training step we sample $t \sim \mathcal{U}(\epsilon, 1-\epsilon)$, derive the per-token mask probability from a fixed noise schedule, and draw a binary mask $m \in \{0,1\}^T$ over the speech positions to form $x_t$ ($x_{t,i} = y_i$ if $m_i = 0$, else $[\textsc{m}]$).
Following Fast-dLLM~v2~\cite{wu2025fastdllmv2efficientblockdiffusion}, we add the complementary view $\bar{m} = 1 - m$ as a second sample in the same batch, so every position is supervised under both masked and unmasked contexts.

\paragraph{Token-Shift Denoising Loss}
We adopt a next-token prediction parameterization following Fast-dLLM~v2~\cite{wu2025fastdllmv2efficientblockdiffusion}: a masked position $i$ is predicted from the hidden state at position $i\!-\!1$ rather than from the mask token itself.
This shifted-label form preserves the backbone's autoregressive interface while still allowing bidirectional context within each block.
The per-token cross-entropy loss at a masked position $i$ is
\begin{equation}
\ell_i = -\log p_\theta\!\left(y_i \mid \mathbf{c}, x^{(<b(i))}, x_t^{(b(i))}\right),
\label{eq:per_token_loss}
\end{equation}
where $y_i$ is the clean target and $b(i)$ is the block containing $i$; the prediction is conditioned on $\mathbf{c}$, the clean preceding blocks $x^{(<b(i))}$, and the noised current block $x_t^{(b(i))}$.
With $\mathcal{M}_b = \{i \mid x_{t,i} = [\textsc{m}]\}$ the masked positions in block $b$ (excluding padding), the sample-level loss averages per-token losses within each block and then over blocks,
\begin{equation}
\mathcal{L}_{\text{denoise}} = \frac{1}{B} \sum_{b=1}^{B} \frac{1}{|\mathcal{M}_b|} \sum_{i \in \mathcal{M}_b} \ell_i.
\label{eq:denoise}
\end{equation}

\subsection{Inference}
\label{sec:inference}

\subsubsection{Block-Autoregressive Decoding}
\label{sec:inference_blockdec}

Block-diffusion inference commits the sequence one block at a time in left-to-right order, with masked positions in the current block unmasked in parallel.
Already-committed blocks act as clean context, and their key-value caches are appended sequentially to provide the inter-block context for subsequent blocks.
Since the conditioning prefix is encoded causally and never attends to speech tokens, its key-value cache depends only on $\mathbf{c}$ and is computed once at the start of inference and reused across every block.
At each step, the forward pass therefore operates only on the current block, accessing the prefix and previously committed blocks through their appended caches.

At each step within a block, the forward pass produces a predictive distribution $p_i^{(k)}$ and a predicted token $\hat{x}_i^{(k)} = \arg\max_v p_i^{(k)}(v)$ at every masked position~$i$.
Two decisions must then be made: \emph{which} positions to commit at this step (\Cref{sec:inference_pmi}) and \emph{how many} positions to commit (\Cref{sec:unmasking_schedule}).
When classifier-free guidance is used (\Cref{sec:inference_cfg}), $p_i^{(k)}$ refers to the conditional branch alone, while a separate guidance combination determines $\hat{x}_i^{(k)}$.

\subsubsection{Prior-Calibrated Scoring}
\label{sec:inference_pmi}

Block-level commitment can induce \emph{boundary-induced context truncation} (BICT)~\cite{shu2026deferred}: once a few positions in a block are committed to incorrect tokens, subsequent blocks decode on top of corrupted context.
This is particularly pronounced for discrete speech codecs, where strong local acoustic dependencies between adjacent frames coexist with a small set of \emph{dominant tokens}---such as silence or low-energy frames---that occupy a disproportionate share of the marginal distribution~\cite{compressed2fine2025, whereunmask2026}; mistakenly committing such a token near a block boundary breaks acoustic continuity.

To address this, we propose \emph{prior-calibrated scoring}, which assigns each masked position an ordering score for parallel unmasking.
The common choice is the model's per-position confidence $p_i^{(k)}(\hat{x}_i^{(k)})$, but for discrete speech codecs this tends to assign large values to dominant tokens regardless of context, causing them to be unmasked preferentially.
To separate this marginal bias from the contextual prediction, we use a pointwise mutual information (PMI) score,
\begin{equation}
\label{eq:pmi}
s_i^{(k)} = \log p_i^{(k)}(\hat{x}_i^{(k)}) - \log \bar{p}(\hat{x}_i^{(k)}),
\end{equation}
where $\bar{p}$ is a reference distribution measuring the marginal probability of producing $\hat{x}_i^{(k)}$ irrespective of local context.
The first term is the model's log-confidence at position $i$, and the second subtracts the marginal share of the same token under $\bar{p}$, so that $s_i^{(k)}$ measures how specifically the predicted token is licensed by local context.

A natural choice for $\bar{p}$ is the in-block marginal averaged over the current block's predictive distributions, but this prior is itself shaped by $\mathbf{c}$, making the score partially self-referential.
We instead use the \emph{unconditional block prior}, computed once from a single forward pass on all-masked sequence $[\textsc{m}]^D$ with conditioning embeddings zeroed,
\begin{equation}
\label{eq:uncond_prior}
\bar{p}(v) = \frac{1}{D} \sum_{j=1}^{D} p_\theta\!\left(v \,\middle|\, [\textsc{m}]^D,\, \mathbf{c} = \mathbf{0}\right)_{\!j}.
\end{equation}
Since $\bar{p}$ depends only on $(D, \theta)$, it is cached for the lifetime of the model.

\subsubsection{Unmasking Schedule}
\label{sec:unmasking_schedule}
Beyond which positions to unmask, the number of positions unmasked at each step must also be chosen.
Committing too many positions in a single step risks introducing incorrect tokens as misleading context, while committing too few requires nearly all $K$ steps and increases inference cost.

\paragraph{Time-Shifted Schedule}
LaViDa~\cite{li2026lavida} introduced a time-shifted (TS) schedule that biases unmasking away from uniform, also adopted in OmniVoice~\cite{zhu2026omnivoice} as its decoding rule: at each step, the TS schedule determines \emph{how many} positions to unmask, and the model commits the corresponding number of top-confidence positions.
We build on the same TS schedule for time allocation, replacing only the position-selection criterion with prior-calibrated scoring (\Cref{sec:inference_pmi}).
The cumulative fraction of unmasked tokens at step $k$ follows
\begin{equation}
\label{eq:rk}
r_k = \frac{\tau \cdot (k/K)}{1 + (\tau - 1) \cdot (k/K)},
\end{equation}
with $r_0 = 0$, total steps $K$, and shift parameter $\tau$; the fraction of newly unmasked tokens at step $k$ is $f_k = r_k - r_{k-1}$.
The optimal $\tau$ differs across configurations such as $K$ and the target token distribution, and we explored different $\tau$ values in our experiments.

\paragraph{Early Decoding}
\label{par:inference_early}
We further adapt the unmasking fraction at each step based on the prior-calibrated scores $s_i^{(k)}$ of \Cref{eq:pmi}.
At step $k$, positions whose score exceeds a threshold $\theta_k$ are unmasked,
\begin{equation}
\label{eq:theta_k}
\theta_k = \mathrm{Quantile}\!\bigl(\{s_i^{(k)}\}_{i \in \mathcal{M}},\, q_k\bigr),
\end{equation}
\begin{equation}
\label{eq:q_k}
q_k = \max\!\left(0,\, 1 - \alpha \cdot \tfrac{k+1}{K}\right),
\end{equation}
where $\mathcal{M}$ is the set of masked positions in the current block and $\alpha \in [0, 1]$ controls how quickly the threshold relaxes.
At early steps, $q_k$ is close to one and only a few positions in the upper region are unmasked; as $k$ increases, $q_k$ decreases and more positions become eligible.
Denoting the resulting fraction by $g_k = q_{k-1} - q_k$, we combine the two schedules via $\max(f_k, g_k)$, so that the TS schedule provides per-step commits while early decoding adds further commits when calibrated confidence is high.
Decoding terminates when all positions in the block have been unmasked or when $K$ steps have been reached; under large $\alpha$, the average step count drops below $K$.
Following OmniVoice~\cite{zhu2026omnivoice}, we additionally support sampling temperatures at both the token and position levels, the latter realized as Gumbel-perturbed selection over the prior-calibrated scores with temperature $\beta = 5$; their effect is studied in \Cref{sec:ablation}.

\subsubsection{Classifier-Free Guidance}
\label{sec:inference_cfg}
We combine classifier-free guidance (CFG)~\cite{ho2022classifierfreediffusionguidance} with prior-calibrated decoding by running both a conditional and an unconditional forward at each step, the latter with the conditioning embeddings replaced by zero vectors, yielding logits $\ell_i^c$ and $\ell_i^u$.
Token sampling uses the standard combination $\ell_i = (1 + w) \ell_i^c - w \ell_i^u$ to determine the predicted token $\hat{x}_i^{(k)} = \arg\max_v \mathrm{softmax}(\ell_i)_v$, while the prior-calibrated score is evaluated on the conditional branch alone (i.e., $p_i^{(k)} = \mathrm{softmax}(\ell_i^c)$ in \Cref{eq:pmi}).
This decoupling---CFG-guided sampling for the committed token, conditional-only PMI for position ranking---keeps the ranking insensitive to $w$ while still routing the CFG-guided token through \Cref{eq:pmi}.
We use $w = 1.0$ by default; a full sweep of $w$ and an alternative mode combining $s_i^c$ and $s_i^u$ are provided in \Cref{app:cfg}.

\begin{table*}[!t]
\centering
\small
\setlength{\tabcolsep}{5pt}
\renewcommand{\arraystretch}{1.05}
\begin{tabular}{l c c | c c c | c c c}
\toprule
\multirow{2}{*}{\textbf{Model}} & \multirow{2}{*}{\textbf{\#Params}} & \multirow{2}{*}{\textbf{Steps}}
& \multicolumn{3}{c|}{\textbf{LibriSpeech-PC test-clean}} & \multicolumn{3}{c}{\textbf{Seed-TTS test-en}} \\
\cmidrule(lr){4-6}\cmidrule(lr){7-9}
& & & \textbf{SIM-o}$\uparrow$ & \textbf{WER}$\downarrow$ & \textbf{UTMOS}$\uparrow$
    & \textbf{SIM-o}$\uparrow$ & \textbf{WER}$\downarrow$ & \textbf{UTMOS}$\uparrow$ \\
\midrule
Ground-truth & -- & -- & 0.690 & 1.87 & 4.10 & 0.734 & 2.14 & 3.52 \\
\midrule
\multicolumn{9}{l}{\textit{Autoregressive Models}} \\
\midrule
IndexTTS2~\cite{zhou2026indextts2}$^\ddagger$    & 1.7B & -- & 0.700 & 2.35 & 4.06 & 0.706 & 2.33 & 3.65 \\
CosyVoice3~\cite{du2025cosyvoice}$^\ddagger$     & 1.1B & -- & 0.694 & \textbf{1.59} & 4.28 & 0.696 & 2.17 & 3.96 \\
VoxCPM~\cite{zhou2025voxcpm}$^\ddagger$          & 0.7B & -- & \textbf{0.717} & 1.74 & 4.18 & \textbf{0.731} & 1.92 & 3.77 \\
Qwen3-TTS~\cite{hu2026qwen3}$^\ddagger$          & 1.1B & -- & 0.704 & 1.60 & \textbf{4.41} & 0.708 & \textbf{1.54} & \textbf{4.16} \\
Chatterbox~\cite{chatterboxtts2025}$^\ddagger$   & 0.5B & -- & 0.707 & 1.99 & 4.29 & 0.685 & 2.20 & 4.10 \\
\midrule
\multicolumn{9}{l}{\textit{Non-Autoregressive Models}} \\
\midrule
F5-TTS~\cite{chen2025f5}$^\ddagger$                 & 0.4B & -- & 0.655 & 1.89 & 3.89 & 0.664 & 1.85 & 3.72 \\
ZipVoice~\cite{zhu2025zipvoice}$^\ddagger$          & 0.1B & -- & 0.668 & 1.64 & 3.98 & 0.697 & 1.70 & 3.82 \\
MaskGCT~\cite{wang2025maskgct}$^\ddagger$           & 2.2B & -- & 0.691 & 2.26 & 3.91 & 0.713 & 2.88 & 3.55 \\
OmniVoice-Emilia~\cite{zhu2026omnivoice}$^\ddagger$ & 0.8B & -- & 0.697 & 1.57 & 4.23 & 0.717 & 1.72 & 3.88 \\
OmniVoice~\cite{zhu2026omnivoice}$^\ddagger$        & 0.8B & -- & \textbf{0.729} & \textbf{1.30} & 4.28 & \textbf{0.741} & \textbf{1.60} & 3.91 \\
\midrule
\multicolumn{9}{l}{\textit{Block-Autoregressive Models (Chatterbox-Flash, Ours)}} \\
\midrule
\quad w/ Fast-dLLM~v2 decoding$^\dagger$          & 0.5B & 10  & 0.656 & 15.36 & 4.14 & 0.646 & 14.49 & 4.00 \\
\quad w/ TS schedule                              & 0.5B & 8   & 0.714 & 1.69  & \textbf{4.29} & 0.703 & 1.97 & \textbf{4.09} \\
\quad w/ TS schedule + ED ($\alpha = 0.5$)        & 0.5B & 5.55 & 0.703 & 1.83 & 4.28 & 0.693 & 2.20 & 4.08 \\
\quad w/ \textbf{PMI} ($\alpha = 0$, ours)        & 0.5B & 8   & \textbf{0.717} & \textbf{1.67} & \textbf{4.29} & \textbf{0.704} & \textbf{1.96} & \textbf{4.09} \\
\quad w/ \textbf{PMI + ED} ($\alpha = 0.5$, ours) & 0.5B & 6.4 & 0.713 & 1.67 & 4.28 & 0.704 & 2.04 & 4.08 \\
\bottomrule
\end{tabular}
\caption{Objective evaluation on zero-shot TTS benchmarks under \textbf{offline} decoding. Best results per AR / NAR / Block-AR group are in \textbf{bold}. Rows marked $\ddagger$ are reproduced from OmniVoice~\cite{zhu2026omnivoice}; ours use the same test sets and metrics (\Cref{sec:exp_setup}), so all columns are comparable. \textbf{Steps} is the average denoising steps per block (Chatterbox-Flash only). Block-AR variants share one configuration ($D{=}16$, $\tau{=}0.5$, $w{=}1.0$, $T{=}0.2$, $\beta{=}5$), differing only in decoding: Fast-dLLM~v2 ($K{=}10$, threshold $0.3$, top-$p$ $0.95$, batch $16$), TS schedule ($K{=}8$), PMI ($K{=}8$), and both with early decoding (+ED, $\alpha{=}0.5$).}
\label{tab:main_results}
\end{table*}

\section{Experiments}
\label{sec:experiments}



\subsection{Experimental Setup}
\label{sec:exp_setup}

\paragraph{Evaluation Benchmarks}
We evaluate on two publicly available English zero-shot TTS benchmarks.
LibriSpeech-PC test-clean is a zero-shot voice cloning benchmark built on the test-clean split of LibriSpeech-PC~\cite{meister2023librispeechpcbenchmarkevaluationpunctuation}, and Seed-TTS test-en is the English evaluation set introduced by Seed-TTS~\cite{anastassiou2406seed}.

\paragraph{Metrics}
We report SIM-o (cosine similarity between WavLM-ECAPA-TDNN~\cite{desplanques20_interspeech, Chen_2022} speaker embeddings of generated vs.\ reference speech), WER (HuBERT~\cite{hsu2021hubertselfsupervisedspeechrepresentation} ASR transcription on LibriSpeech-PC, Whisper-large-v3~\cite{radford2022robustspeechrecognitionlargescale} on Seed-TTS test-en, against the input text), and UTMOS~\cite{saeki2022utmosutokyosarulabvoicemoschallenge} for naturalness.
Metric configuration and baseline numbers are taken from OmniVoice~\cite{zhu2026omnivoice}.

\paragraph{Implementation Details}
The model replaces the T3 decoder of Chatterbox-TTS with the block-diffusion architecture described in \Cref{sec:method}, trained with the hybrid attention mask and the token-shift loss.
We initialize from a pretrained Chatterbox-TTS checkpoint and continue training with AdamW using a cosine learning rate schedule (peak $10^{-5}$, $10\%$ warmup) with an effective batch size of $440$ in bfloat16 precision.
The model is trained with a block size of $D = 32$.
Text inputs are normalized by a custom preprocessor that converts numbers, dates, times, and other non-standard tokens into their spoken forms before tokenization.
At inference time, our canonical configuration is block size $D = 16$, number of denoising steps $K = 8$, TS schedule parameter $\tau = 0.5$, CFG scale $w = 1.0$, sampling temperature $0.2$, and position temperature $\beta = 5$.
We report two main settings differing only in the early-decoding parameter $\alpha$: a quality-strongest setting at $\alpha = 0$ (no early decoding) and an efficiency-oriented setting at $\alpha = 0.5$.
All experiments are conducted on NVIDIA H100 GPUs, and inference uses attention kernels and paged key-value cache management from FlashInfer~\cite{ye2025flashinferefficientcustomizableattention}; further implementation details are provided in \Cref{app:impl}.

\subsection{Main Results}
\label{sec:results}

\Cref{tab:main_results} reports zero-shot TTS performance against recent state-of-the-art autoregressive (AR) and non-autoregressive (NAR) baselines, under the canonical configuration without early decoding ($\alpha = 0$, $K = 8$).

Against Chatterbox, the AR backbone we build on, the AR-to-block-diffusion conversion improves both SIM-o and WER on the two benchmarks while leaving UTMOS unchanged, unlocking parallel decoding at preserved or improved perceptual quality.
Within the NAR group, Chatterbox-Flash trails only OmniVoice on LibriSpeech-PC SIM-o and WER despite training on $70$k hours of English versus OmniVoice's $581$k hours of multilingual data, attains the best NAR UTMOS on LibriSpeech-PC, and surpasses F5-TTS, MaskGCT, and OmniVoice-Emilia on WER. It is also the only NAR system here with native streaming support (\Cref{app:streaming_baselines}).

\begin{table*}[!t]
\centering
\small
\setlength{\tabcolsep}{6pt}
\renewcommand{\arraystretch}{1.05}
\begin{tabular}{l l c c c | c c c}
\toprule
\multirow{2}{*}{\textbf{Mode}} & \multirow{2}{*}{\textbf{Method}}
& \multicolumn{3}{c|}{\textbf{LibriSpeech-PC test-clean}}
& \multicolumn{3}{c}{\textbf{Seed-TTS test-en}} \\
\cmidrule(lr){3-5}\cmidrule(lr){6-8}
& & \textbf{SIM-o}$\uparrow$ & \textbf{WER}$\downarrow$ & \textbf{UTMOS}$\uparrow$
  & \textbf{SIM-o}$\uparrow$ & \textbf{WER}$\downarrow$ & \textbf{UTMOS}$\uparrow$ \\
\midrule
\multirow{4}{*}{\textbf{Offline}}
& TS schedule ($\alpha{=}0$)   & 0.714 & 1.69 & \textbf{4.29} & 0.703 & 1.97 & \textbf{4.09} \\
& TS schedule ($\alpha{=}0.5$) & 0.703 & 1.83 & 4.28          & 0.693 & 2.20 & 4.08 \\
& PMI ($\alpha{=}0$)           & \textbf{0.717} & \textbf{1.67} & \textbf{4.29} & \textbf{0.704} & 1.96 & \textbf{4.09} \\
& PMI ($\alpha{=}0.5$)         & 0.713 & \textbf{1.67} & 4.28  & \textbf{0.704} & 2.04 & 4.08 \\
\midrule
\multirow{4}{*}{\textbf{Streaming}}
& TS schedule ($\alpha{=}0$)   & \textbf{0.703} & \textbf{1.66}          & 4.19 & 0.687 & 1.79          & 3.99 \\
& TS schedule ($\alpha{=}0.5$) & 0.702 & 1.76          & 4.19 & 0.689 & 1.75 & 3.99 \\
& PMI ($\alpha{=}0$)           & \textbf{0.703} & 1.70 & \textbf{4.20} & \textbf{0.690} & \textbf{1.73} & 3.99 \\
& PMI ($\alpha{=}0.5$)         & 0.702 & 1.80  & \textbf{4.20} & \textbf{0.690} & 1.96 & \textbf{4.00} \\
\bottomrule
\end{tabular}
\caption{Offline vs streaming quality on the two standard benchmarks under the canonical configuration. Best per column within each mode in \textbf{bold}. Offline rows match the Block-AR group of \Cref{tab:main_results}.}
\label{tab:off_str_quality}
\end{table*}

\begin{table*}[!t]
\centering
\small
\setlength{\tabcolsep}{5pt}
\renewcommand{\arraystretch}{1.05}
\begin{tabular}{l l cc cc cc cc}
\toprule
\multirow{2}{*}{\textbf{Mode}} & \multirow{2}{*}{\textbf{Method}}
& \multicolumn{2}{c}{\textbf{Overall WER}$\downarrow$}
& \multicolumn{2}{c}{\textbf{Pronunciation}$\downarrow$}
& \multicolumn{2}{c}{\textbf{Paraling.}$\downarrow$}
& \multicolumn{2}{c}{\textbf{Foreign}$\downarrow$} \\
\cmidrule(lr){3-4}\cmidrule(lr){5-6}\cmidrule(lr){7-8}\cmidrule(lr){9-10}
& & Off & Str & Off & Str & Off & Str & Off & Str \\
\midrule
\multirow{2}{*}{TS}
& $\alpha{=}0$   & 37.97 & 36.25 & 73.02 & 71.35 & 27.69 & 23.36 & 17.59 & 18.44 \\
& $\alpha{=}0.5$ & 35.75 & \textcolor{red!65!black}{39.53} & 74.53 & \textcolor{red!65!black}{85.88} & 19.51 & 19.98 & 18.06 & 18.54 \\
\midrule
\multirow{2}{*}{PMI}
& $\alpha{=}0$   & 34.86 & 35.10 & 71.04 & 72.32 & 19.55 & 19.60 & 18.52 & 18.05 \\
& $\alpha{=}0.5$ & \textbf{34.79} & \textbf{35.68} & \textbf{70.28} & \textbf{72.36} & 19.61 & 20.32 & 18.90 & 18.93 \\
\bottomrule
\end{tabular}
\caption{Per-category WER on EmergentTTS-Eval~\cite{manku2026emergenttts}, offline (Off) vs.\ streaming (Str). Best value per column within the PMI/TS $\alpha{=}0.5$ streaming setting in \textbf{bold}; \textcolor{red!65!black}{red} marks the TS baseline's degradation under streaming. Model-as-a-judge MOS is stable across all rows ($3.50$--$3.55$) and omitted. \textit{Pronunciation}: rare or non-standard pronunciation; \textit{Paraling.}: paralinguistic cues; \textit{Foreign}: foreign words.}
\label{tab:emergent_hard}
\end{table*}

The \textit{Block-Autoregressive Models} group compares five decoding configurations under this canonical setting.
Fast-dLLM~v2 decoding transfers poorly to discrete speech codecs (WER $> 14$ on both benchmarks) despite a carefully tuned threshold: its fixed confidence bar fires first on the dominant low-information tokens, locking in a corrupted context for the rest of the block---the boundary-induced truncation that motivates prior calibration (\Cref{sec:inference_pmi}).
The TS schedule alone (top-confidence selection, as in OmniVoice) and PMI ($\alpha = 0$) achieve statistically comparable quality, so PMI's decisive contribution is not raw quality but the calibration of its scores, which serves as a reliable thresholding signal for early decoding.
Isolating this from the scheduling rule, PMI+ED attains lower WER than TS+ED at $\alpha = 0.5$ ($1.67$ vs.\ $1.83$ on LibriSpeech-PC, $2.04$ vs.\ $2.20$ on Seed-TTS) at a comparable step budget, confirming that the calibrated signal, not the schedule alone, drives the quality-preserving step reduction.
PMI+ED matches the no-ED TS baseline's quality while cutting the average step count from $8$ to $\sim$$6.3$/block, a $\sim$$20\%$ reduction.

\subsection{Offline vs Streaming}
\label{sec:off_str}

The results above are offline. Under our block-causal streaming pipeline (\Cref{app:streaming_efficiency}), the two decoding methods stay close and the offline-to-streaming shift is small (\Cref{tab:off_str_quality}), so quality holds under the streaming constraint. These benchmarks saturate near WER $1.7$--$2.0$, however, leaving little room to separate the methods, which motivates the harder evaluation that follows.

\subsection{Challenging TTS Scenarios}
\label{sec:hard}

The standard benchmarks are dominated by read speech, where the model's confidence and the marginal token distribution rarely disagree. Prior calibration is built for harder cases, so we evaluate on EmergentTTS-Eval~\cite{manku2026emergenttts}, which targets complex prosodic, expressive, and linguistically difficult utterances (\Cref{tab:emergent_hard}). At the overall level the asymmetry is stark: moving from offline to streaming raises the TS baseline's WER by $3.78$ ($35.75 \to 39.53$) but PMI's by only $0.89$ ($34.79 \to 35.68$).

The gap is driven by \textit{pronunciation}, the hardest category, where the TS baseline with early decoding jumps to $85.88$ under streaming while PMI stays around $70$--$72$; because this category carries the most word errors, its collapse drives the overall degradation. The other categories show where calibration does not help: \textit{foreign words} depend on lexical coverage rather than position selection, and \textit{paralinguistics} is capped near $19$--$20$ by the backbone's limited handling of non-lexical vocalizations. PMI's advantage is thus real but category-specific---it recovers intelligibility where dominant-token bias corrupts position selection, and is neutral where the bottleneck lies elsewhere.

\begin{table}[!t]
\centering
\small
\setlength{\tabcolsep}{6pt}
\renewcommand{\arraystretch}{1.1}
\begin{tabular}{l c c}
\toprule
\textbf{Metric} & \textbf{ElevenLabs v3} & \textbf{Chatterbox-Flash} \\
\midrule
NMOS mean$\uparrow$            & \textbf{4.04}\,\small{$\pm$0.26} & 3.91\,\small{$\pm$0.23} \\
\quad \% $\leq 2$ $\downarrow$ & 12.9 & \textbf{8.6} \\
\quad \% $\geq 4$ $\uparrow$   & \textbf{80.0} & 67.1 \\
\midrule
SMOS mean$\uparrow$            & 3.50\,\small{$\pm$0.35} & \textbf{4.56}\,\small{$\pm$0.18} \\
\bottomrule
\end{tabular}
\caption{Human evaluation on $10$ utterances from Seed-TTS test-en, $7$ ratings each ($70$ per system), on 5-point Likert scales (NMOS: naturalness; SMOS: speaker similarity). Means are shown with $95\%$ confidence intervals. Chatterbox-Flash is evaluated in its default offline configuration ($D{=}16$, $\alpha{=}0.5$).}
\label{tab:human_eval}
\end{table}

\subsection{Human Evaluation}
\label{sec:human_eval}

We conducted a side-by-side evaluation against ElevenLabs v3, a frontier commercial zero-shot TTS system, collecting naturalness (NMOS) and speaker-similarity (SMOS) ratings on 5-point Likert scales over $10$ Seed-TTS utterances, $7$ ratings each ($70$ per system).
On NMOS (\Cref{tab:human_eval}), Chatterbox-Flash and ElevenLabs v3 attain comparable means ($3.91$ vs.\ $4.04$), and the two confidence intervals overlap substantially ($\pm0.23$ vs.\ $\pm0.26$), so mean naturalness is statistically indistinguishable. The difference instead shows in the tails: Chatterbox-Flash produces fewer MOS-$\leq 2$ ratings ($8.6\%$ vs.\ $12.9\%$), indicating fewer catastrophic failures even when the means match. On SMOS the means favor Chatterbox-Flash ($4.56$ vs.\ $3.50$) and the intervals do not overlap ($\pm0.18$ vs.\ $\pm0.35$), so the speaker-similarity advantage is a reliable effect rather than sampling noise.

\subsection{Ablation Studies}
\label{sec:ablation}

We ablate two inference-time hyperparameters affecting the quality--compute trade-off at the canonical configuration: the block size $D$ and the denoising step budget $K$ in combination with the early-decoding parameter $\alpha$.

\begin{figure}[!t]
\centering
\includegraphics[width=\columnwidth]{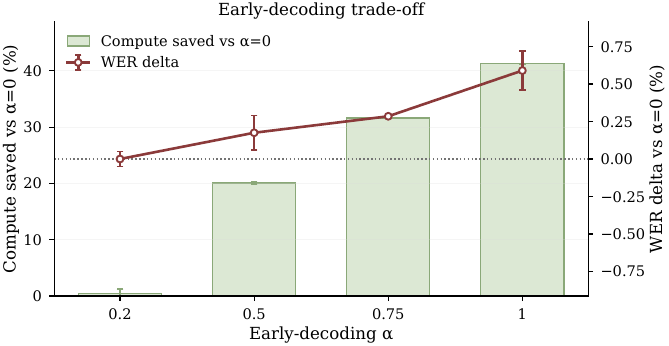}
\caption{Early-decoding trade-off at $D = 16$, $K = 10$, averaged across LibriSpeech-PC and Seed-TTS test-en. Bars: number of computing steps saved relative to $\alpha = 0$. Line: WER delta versus $\alpha = 0$, with error bars spanning the two benchmarks. At $\alpha = 0.5$, $\sim$$20\%$ steps are saved at negligible WER cost; further savings up to $\sim$$41\%$ at $\alpha = 1$ incur a $\sim$$+0.6$ WER increase.}
\label{fig:tau_tradeoff}
\end{figure}

\begin{figure*}[!t]
\centering
\includegraphics[width=0.95\textwidth]{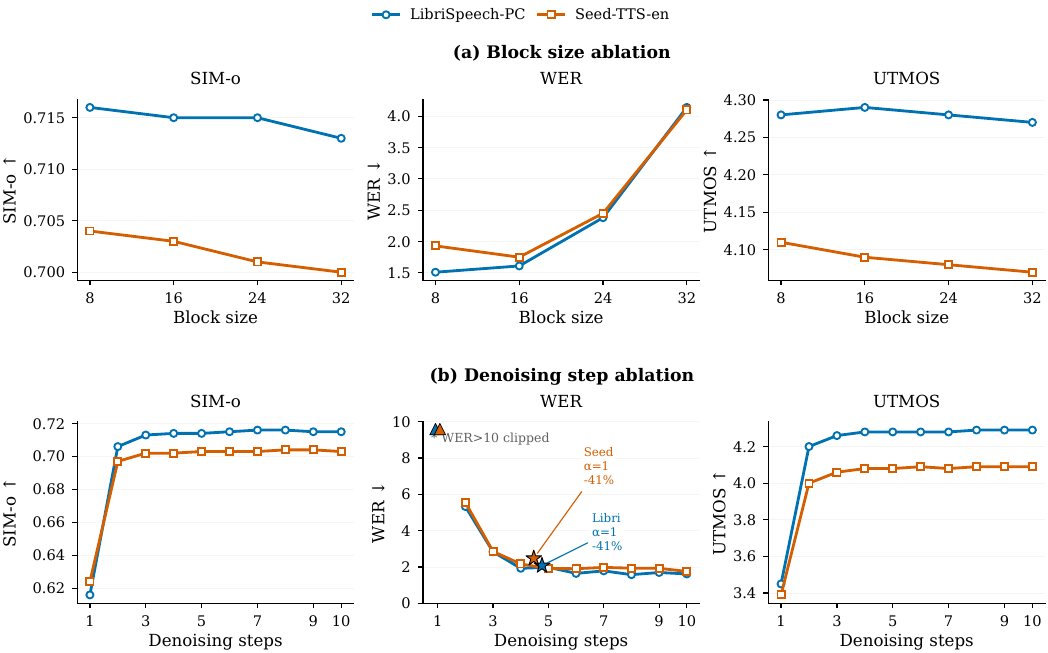}
\caption{Inference-time ablations on LibriSpeech-PC (blue, circles) and Seed-TTS test-en (orange, squares). \textbf{(a) Block size $D$} and \textbf{(b) denoising step budget $K$} (fixed-step, $\alpha{=}0$). Triangles: WER off the display range ($K{\leq}2$). Stars: adaptive early decoding ($\alpha{=}1$), reaching plateau-level WER at $\sim$$4.6$ steps/block. Trends discussed in \Cref{sec:ablation}.}
\label{fig:block_steps}
\end{figure*}

\paragraph{Block Size}
\Cref{fig:block_steps}(a) sweeps the inference block size $D \in \{8, 16, 24, 32\}$ with the model trained at $D = 32$.
SIM-o and UTMOS are essentially flat across the range, and WER stays within noise up to $D = 16$ before degrading sharply at $D \geq 24$.
This degradation reflects the difficulty of parallel unmasking on large blocks: these configurations require committing more positions per step than the model can confidently rank, even with prior calibration.

\paragraph{Step Budget}
\Cref{fig:block_steps}(b) characterizes the effect of step size $K$ in fixed-step decoding at $D = 16$.
$K \leq 2$ cannot recover the speech sequence, but quality stabilizes rapidly and plateaus by $K \geq 6$.
We adopt $K = 8$ as the default, providing modest headroom over the plateau for adaptive early termination.
The stars mark adaptive early decoding at $\alpha = 1$, which reaches the same plateau-level WER at $\sim$$4.6$ average steps without exceeding the fixed-step plateau.

\paragraph{Early Decoding}
\Cref{fig:tau_tradeoff} sweeps the early-decoding parameter $\alpha \in \{0.2, 0.5, 0.75, 1.0\}$ at $D = 16$, $K = 10$.
At $\alpha = 0.2$ early decoding is almost inactive, but from $\alpha = 0.5$ onward it meaningfully tightens the step budget while the WER cost grows only mildly, staying within $+0.6$ of the no-ED baseline even at $\alpha = 1.0$.
We adopt $\alpha = 0.5$ as the canonical efficiency-oriented setting ($\sim$$20\%$ compute reduction at negligible quality cost); $\alpha = 0.75$--$1.0$ offer further savings for latency-critical deployments.

\paragraph{Further Ablations}
Sweeps of the CFG scale $w$, sampling temperature $T$, and position temperature $\beta$, together with a head-to-head comparison of the TS schedule baseline against PMI across step budgets $K \in \{2, 5, 8\}$, are provided in \Cref{app:ablations}.
The comparison shows that PMI and the TS schedule baseline track each other closely across the step range, confirming that the choice of decoding method contributes only marginally to quality at any compute budget---PMI's advantage lies in providing the calibrated confidence signal that enables adaptive early decoding.

\section{Conclusion}
\label{sec:conclusion}

We presented Chatterbox-Flash, a block-diffusion zero-shot TTS model obtained by fine-tuning a pretrained autoregressive decoder into a parallel masked decoder while preserving block-by-block streaming.
Our key contributions are an inference-time prior-calibrated scoring scheme, which suppresses the dominant-token bias of discrete speech codecs and provides a well-calibrated confidence signal; an early-decoding schedule that adaptively terminates iteration; and a streaming-compatible inference engine that combines the two.
On standard zero-shot TTS benchmarks, it matches strong AR and NAR baselines in quality despite training on substantially less data, achieves the highest UTMOS among NAR baselines on both LibriSpeech-PC and Seed-TTS, and reduces the average denoising step count by $\sim$$20\%$ at minimal quality cost.
While prior calibration is largely interchangeable with a top-confidence baseline on saturated read-speech benchmarks, its value emerges in the regime the model is built for: on challenging inputs decoded under streaming, it keeps word error low where the baseline degrades sharply.
These results indicate that block diffusion with calibration-aware inference is a viable design point for production-grade streaming zero-shot TTS.

\section*{Limitations}

We prioritized training stability over data-related ablations, leaving the contribution of individual data sources unisolated, and the model still collapses at substantially larger block sizes ($D \geq 128$) while OmniVoice's full-sequence formulation remains stable; alternative recipes (\Cref{app:block_scaling}) reach larger $D$ only under restricted conditions, leaving the parallelism gap with full-sequence models open.
Finally, at saturated compute PMI and the TS schedule baseline yield statistically equivalent objective metrics (\Cref{tab:beta_ablation}), so PMI's advantage there is the calibrated confidence it provides for early decoding rather than a direct quality gain; on harder out-of-distribution inputs it does reduce WER directly, but whether this holds under tighter budgets or out-of-domain reference speakers is left to future work.

\bibliography{custom}

\newpage

\appendix

\section{Attention Kernel Implementation}
\label{app:attn_kernel}

The hybrid attention mask described in \Cref{sec:training} is implemented using either (i)~PyTorch's native \texttt{flex\_attention} with a custom mask function, or (ii)~MagiAttention's Flex-Flash-Attention (FFA)~\cite{ai2025magi1autoregressivevideogeneration, magiattention2025}\footnote{\url{https://github.com/SandAI-org/MagiAttention}} kernel for higher throughput on long sequences.

\begin{table*}[!h]
\centering
\small
\setlength{\tabcolsep}{8pt}
\renewcommand{\arraystretch}{1.05}
\begin{tabular}{c c c c c c c}
\toprule
\multirow{2}{*}{\textbf{$T$}}
& \multicolumn{3}{c}{\textbf{LibriSpeech-PC test-clean}} & \multicolumn{3}{c}{\textbf{Seed-TTS test-en}} \\
\cmidrule(lr){2-4} \cmidrule(lr){5-7}
& \textbf{SIM-o}$\uparrow$ & \textbf{WER}$\downarrow$ & \textbf{UTMOS}$\uparrow$
& \textbf{SIM-o}$\uparrow$ & \textbf{WER}$\downarrow$ & \textbf{UTMOS}$\uparrow$ \\
\midrule
0.2 (default) & 0.715 & \textbf{1.61} & 4.29 & 0.703 & \textbf{1.75} & 4.09 \\
0.4           & 0.715 & 1.64 & 4.29 & 0.704 & 1.82 & 4.08 \\
0.6           & 0.714 & 1.75 & 4.28 & 0.702 & 1.93 & 4.07 \\
0.8           & 0.713 & 1.75 & 4.26 & 0.702 & 2.05 & 4.05 \\
1.0           & 0.714 & 2.01 & 4.24 & 0.702 & 2.15 & 4.02 \\
\bottomrule
\end{tabular}
\caption{Sampling temperature $T$ sweep at $\beta = 5$, others at canonical default.}
\label{tab:temperature_ablation}
\end{table*}

\begin{figure}[t]
    \centering
    \includegraphics[width=\columnwidth]{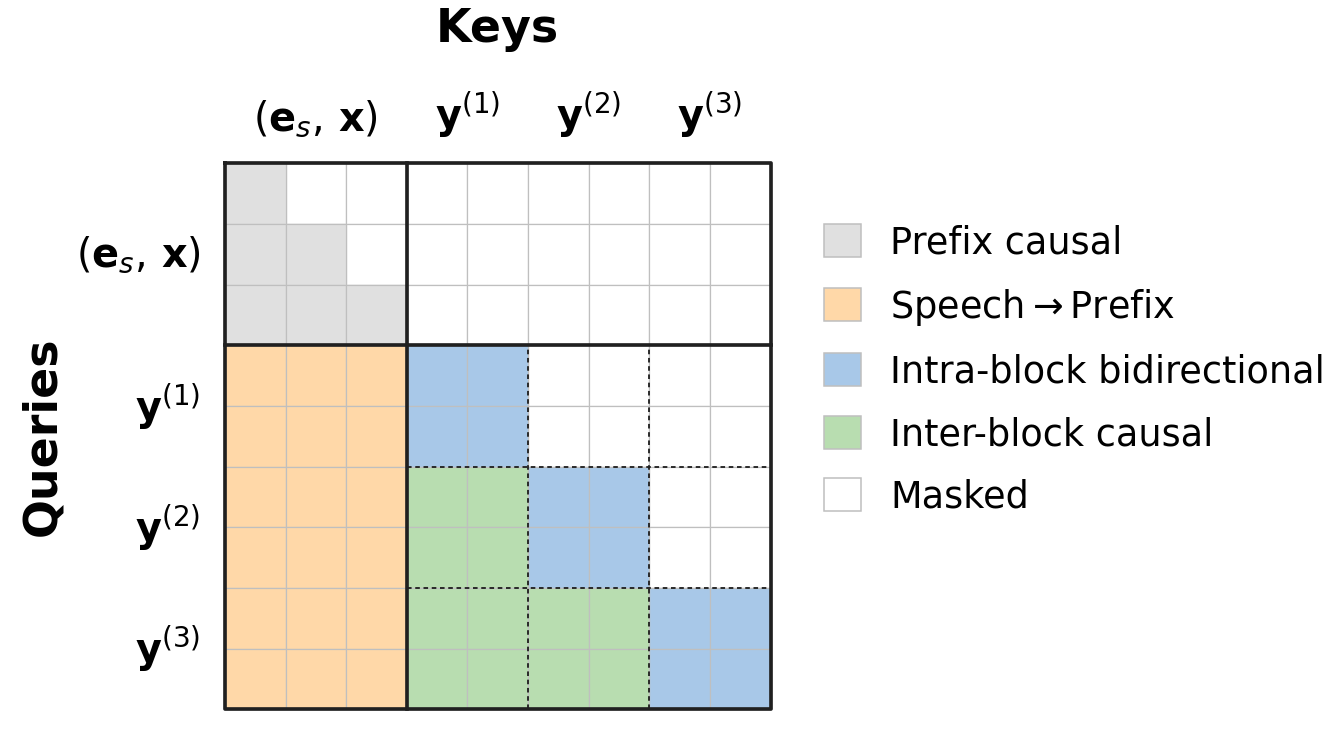}
    \caption{Hybrid block attention mask at $D=2$. A colored cell at row $i$, column $j$ indicates that query position~$i$ attends to key position~$j$. The conditioning prefix $(\mathbf{e}_s, \mathbf{x})$ uses causal attention (gray); speech tokens attend to the entire prefix (orange); within each speech block $\mathbf{y}^{(b)}$, attention is bidirectional (blue); across blocks, attention is causal (green). Future blocks are never visible to past blocks, enabling block-by-block streaming.}
    \label{fig:attn_mask}
\end{figure}

\paragraph{Range Encoding for FFA}
FFA expresses sparse attention as a list of rectangular query/key index ranges with per-range attention types (\texttt{FULL} or \texttt{CAUSAL}).
Letting $L_{\text{pre}} = |\mathbf{e}_s| + |\mathbf{x}|$ denote the prefix length, $N$ the speech-stream length, and $D$ the block size, our hybrid mask decomposes into four rectangle groups:
\begin{enumerate}
    \item \textbf{Prefix Self-Attention} (\texttt{CAUSAL}): $q, k \in [0,\, L_{\text{pre}})$.
    \item \textbf{Speech-to-Prefix} (\texttt{FULL}): $q \in [L_{\text{pre}},\, L_{\text{pre}}{+}N)$, $k \in [0,\, L_{\text{pre}})$.
    \item \textbf{Intra-Block Bidirectional} (\texttt{FULL}): for each block $b$, $q, k \in [L_{\text{pre}}{+}bD,\, L_{\text{pre}}{+}(b{+}1)D)$.
    \item \textbf{Inter-Block Left-Context} (\texttt{FULL}): for $b \geq 1$, $q$ in block $b$, $k$ in blocks $0, \ldots, b{-}1$.
\end{enumerate}

\paragraph{Equivalence}
The boolean mask induced by the FFA range union is identical to the reference mask used by \texttt{flex\_attention}; both backends produce numerically equivalent outputs up to bf16 noise.

\begin{table}[!h]
\centering
\small
\begin{tabular}{lr}
\toprule
\textbf{Dataset} & \textbf{\# Samples} \\
\midrule
\multicolumn{2}{l}{\textit{Public}} \\
MLS-English~\cite{pratap2020mls}           & 10.8M \\
Emilia (en, part 1)~\cite{he2024emilia}    & 9.1M  \\
Loquacious~\cite{parcollet2025loquacious}  & 3.9M  \\
GLOBE~\cite{wang2024globe}                 & 582K  \\
LibriTTS-R~\cite{koizumi2023librittsr}     & 375K  \\
HiFi-TTS~\cite{bakhturina2021hifitts}      & 324K  \\
EARS~\cite{richter2024ears}                & 12K   \\
Expresso~\cite{nguyen2023expresso}         & 12K   \\
\midrule
\multicolumn{2}{l}{\textit{Privately collected}} \\
Audiobook                       & 17.7M \\
Podcasts                        & 726K  \\
IVR (VC-augmented)              & 445K  \\
Short-form utterances$^\dagger$ & 292K  \\
Conversational                  & 50K   \\
Stylized speech                 & 62K   \\
\midrule
\multicolumn{2}{l}{\textbf{Total} \hfill \textbf{43.8M ($\sim$70k hours)}} \\
\bottomrule
\end{tabular}
\caption{Training dataset composition. Sample counts are reported after preprocessing and integrity filtering. $^\dagger$Short-form includes phone numbers, names, time expressions, and expressive single-word utterances.}
\label{tab:dataset_details}
\end{table}

\section{Training Dataset Details}
\label{app:dataset_details}

We train on approximately $70$k hours of English speech ($44$M utterances, $528$k speakers), compiled from publicly available corpora---spanning large-scale read speech~\cite{pratap2020mls, he2024emilia, parcollet2025loquacious, koizumi2023librittsr, bakhturina2021hifitts}, expressive and anechoic speech~\cite{nguyen2023expresso, richter2024ears}, and accented English~\cite{wang2024globe}---together with privately collected audiobook, conversational, and short-form utterances (names, numbers, time expressions).
\Cref{tab:dataset_details} lists the datasets used for training.
The total training set contains $43.8$M utterances ($\sim$70k hours) from $528$k speakers, distributed across public and privately collected sources.
The privately collected portion broadens coverage of conversational, expressive, and short-form utterances beyond standard read-speech corpora.

\begin{table*}[!h]
\centering
\small
\setlength{\tabcolsep}{6pt}
\renewcommand{\arraystretch}{1.05}
\begin{tabular}{c c c | c c c | c c c}
\toprule
\multirow{2}{*}{$K$} & \multirow{2}{*}{\textbf{Method}} & \multirow{2}{*}{\boldmath$\beta$}
& \multicolumn{3}{c|}{\textbf{LibriSpeech-PC test-clean}} & \multicolumn{3}{c}{\textbf{Seed-TTS test-en}} \\
\cmidrule(lr){4-6} \cmidrule(lr){7-9}
& & & \textbf{SIM-o}$\uparrow$ & \textbf{WER}$\downarrow$ & \textbf{UTMOS}$\uparrow$
    & \textbf{SIM-o}$\uparrow$ & \textbf{WER}$\downarrow$ & \textbf{UTMOS}$\uparrow$ \\
\midrule
\multirow{4}{*}{$2$}
& TS schedule & 0 & 0.679 & 15.10 & 4.00 & 0.676 & 15.40 & 3.81 \\
& TS schedule & 5 & 0.697 &  8.74 & 4.13 & 0.690 &  9.13 & 3.93 \\
& PMI         & 0 & 0.690 & 11.82 & 4.08 & 0.682 & 13.98 & 3.86 \\
& PMI         & 5 & 0.707 &  \textbf{5.17} & \textbf{4.21} & 0.697 &  \textbf{5.37} & \textbf{4.01} \\
\midrule
\multirow{4}{*}{$5$}
& TS schedule & 0 & 0.714 &  2.52 & 4.26 & 0.702 &  2.50 & 4.08 \\
& TS schedule & 5 & 0.714 &  1.91 & 4.28 & 0.702 &  \textbf{2.13} & 4.09 \\
& PMI         & 0 & 0.715 &  2.36 & 4.27 & 0.703 &  2.46 & 4.07 \\
& PMI         & 5 & 0.713 &  \textbf{1.96} & 4.27 & 0.703 &  2.01 & 4.08 \\
\midrule
\multirow{4}{*}{$8$}
& TS schedule & 0 & 0.714 &  1.87 & 4.28 & 0.703 &  2.07 & 4.09 \\
& TS schedule & 5 & 0.714 &  1.69 & 4.29 & 0.703 &  1.97 & 4.09 \\
& PMI         & 0 & 0.716 &  1.82 & 4.28 & 0.703 &  2.25 & 4.09 \\
& PMI         & 5 & \textbf{0.717} & \textbf{1.67} & \textbf{4.29} & \textbf{0.704} & \textbf{1.96} & 4.09 \\
\bottomrule
\end{tabular}
\caption{Position temperature $\beta$ ablation at $D = 16$, $\tau = 0.5$, $w = 1.0$, $T = 0.2$. $\beta = 5$ (Gumbel-perturbed selection) consistently outperforms $\beta = 0$ (deterministic top-$n_k$) across both decoding methods (TS schedule, PMI) and all step budgets, with the effect most pronounced at low $K$. Best WER per $K$ is in \textbf{bold}.}
\label{tab:beta_ablation}
\end{table*}

\begin{table*}[!h]
\centering
\small
\setlength{\tabcolsep}{8pt}
\renewcommand{\arraystretch}{1.05}
\begin{tabular}{c c c c c c c}
\toprule
\multirow{2}{*}{\textbf{$w$}}
& \multicolumn{3}{c}{\textbf{LibriSpeech-PC test-clean}} & \multicolumn{3}{c}{\textbf{Seed-TTS test-en}} \\
\cmidrule(lr){2-4} \cmidrule(lr){5-7}
& \textbf{SIM-o}$\uparrow$ & \textbf{WER}$\downarrow$ & \textbf{UTMOS}$\uparrow$
& \textbf{SIM-o}$\uparrow$ & \textbf{WER}$\downarrow$ & \textbf{UTMOS}$\uparrow$ \\
\midrule
0.5            & 0.715 & \textbf{1.52} & \textbf{4.30} & 0.700 & 1.94 & \textbf{4.12} \\
1.0 (default)  & 0.715 & 1.63 & 4.29 & \textbf{0.705} & 1.79 & 4.08 \\
1.5            & 0.715 & 1.71 & 4.27 & 0.703 & \textbf{1.77} & 4.05 \\
2.0            & 0.714 & 1.80 & 4.26 & 0.702 & 1.94 & 4.02 \\
\bottomrule
\end{tabular}
\caption{CFG scale $w$ sweep at the canonical configuration. $w = 1.0$ is the default used in our main results.}
\label{tab:cfg_ablation}
\end{table*}

\begin{table}[!t]
\centering
\small
\setlength{\tabcolsep}{5pt}
\renewcommand{\arraystretch}{1.05}
\begin{tabular}{l c c}
\toprule
\textbf{Model / Config} & \textbf{TTFP (ms)}$\downarrow$ & \textbf{RTF}$\downarrow$ \\
\midrule
\multicolumn{3}{l}{\textit{Qwen3-TTS~\cite{hu2026qwen3} (AR, reported)}} \\
25\,Hz, 1.7B & 150 & 0.253 \\
25\,Hz, 0.6B & 138 & 0.234 \\
12\,Hz, 1.7B & 101 & 0.313 \\
12\,Hz, 0.6B & 97  & 0.288 \\
\midrule
\multicolumn{3}{l}{\textit{Chatterbox-Flash (ours)}} \\
$D = 16$, $\alpha = 0.5$ (default) & 118 & 0.107 \\
$D = 16$, $\alpha = 0.75$          & 106 & 0.091 \\
$D = 24$, $\alpha = 0.5$           & 119 & 0.100 \\
$D = 24$, $\alpha = 0.75$          & 105 & 0.084 \\
$D = 32$, $\alpha = 0.5$           & 115 & 0.090 \\
$D = 32$, $\alpha = 0.75$          & \textbf{103} & \textbf{0.076} \\
\bottomrule
\end{tabular}
\caption{Streaming efficiency at concurrency $1$ over $50$ utterances. \textbf{TTFP}: request-to-first-packet wall-clock time; \textbf{RTF}: generation time / audio duration. Chatterbox-Flash is 25\,Hz, 0.5B, measured by us on H100; best values in \textbf{bold}. Qwen3-TTS numbers are reproduced from its technical report under its own, not fully specified, setup---\emph{not} re-measured by us---and serve only as a loose reference, not a controlled comparison (see Comparability Caveat). Differences in GPU, precision, batching, model size, and token rate mean the two blocks are not directly comparable.}
\label{tab:streaming_efficiency_appendix}
\end{table}

\section{Additional Ablations}
\label{app:ablations}

This appendix reports additional ablations supplementing \Cref{sec:ablation}: sampling and position temperatures (\Cref{app:temperature}) and the CFG scale $w$ (\Cref{app:cfg}).
All sweeps use the canonical configuration of \Cref{sec:exp_setup} except for the swept parameter.

\subsection{Sampling and Position Temperatures}
\label{app:temperature}

\paragraph{Sampling Temperature}
\Cref{tab:temperature_ablation} sweeps the speech-token sampling temperature $T$ at the canonical configuration ($D = 16$, $K = 10$, $\alpha = 0$, $w = 1.0$, $\beta = 5$).
$T = 0.2$ minimizes mean WER on both benchmarks ($1.61$ on LibriSpeech-PC, $1.75$ on Seed-TTS) and is our default; increasing $T$ progressively degrades WER ($+0.4$ at $T = 1.0$) while leaving SIM-o and UTMOS within noise.

\paragraph{Position Temperature and Decoding Method}
For the position temperature $\beta$ (Gumbel perturbation on the prior-calibrated scores; \Cref{sec:unmasking_schedule}), we compared $\beta = 0$ (deterministic top-$n_k$) against $\beta = 5$ across both decoding methods (TS schedule and PMI) and step budgets $K \in \{2, 5, 8\}$ (\Cref{tab:beta_ablation}).
This sweep simultaneously serves as the head-to-head comparison between the TS schedule baseline and PMI referenced in \Cref{sec:ablation}.
$\beta = 5$ consistently lowers WER over $\beta = 0$ on both benchmarks across all settings, with the effect most pronounced at low $K$ (e.g., at $K = 2$: PMI WER drops from $11.82$ to $5.17$ on LibriSpeech-PC, $13.98$ to $5.37$ on Seed-TTS); the same trend holds for the TS schedule.
Comparing the TS schedule baseline and PMI at $\beta = 5$, the two methods track each other closely across all step budgets---PMI is slightly ahead at $K = 5, 8$ on LibriSpeech-PC and at $K = 2$ on both benchmarks, while the TS schedule is slightly ahead at $K = 5$ on Seed-TTS---suggesting that, on these read-speech benchmarks where WER saturates near $1.7$--$2.0$, the choice of decoding method contributes only marginally to quality.
PMI's empirical advantage manifests in two regimes beyond this saturated setting: when early decoding is applied (e.g., $\alpha = 0.5$ in \Cref{tab:main_results}), where the calibrated confidence signal enables the per-step quantile threshold of \Cref{eq:theta_k} to terminate blocks early without quality regression; and on harder, out-of-distribution inputs (\Cref{tab:emergent_hard}), where prior calibration directly reduces WER.
SIM-o and UTMOS also improve modestly with $\beta = 5$.
We adopt $\beta = 5$ throughout.

\subsection{Classifier-Free Guidance Scale}
\label{app:cfg}

\Cref{tab:cfg_ablation} sweeps the CFG scale $w$ at the canonical configuration ($D = 16$, $K = 10$, $\alpha = 0$, $T = 0.2$, $\beta = 5$).
The two benchmarks favor slightly different operating points: LibriSpeech-PC attains its lowest WER at $w = 0.5$ ($1.52$), while Seed-TTS test-en attains its lowest at $w = 1.5$ ($1.77$).
$w = 1.0$ provides the most balanced trade-off across both---$1.63$ / $1.79$ WER with the highest Seed-TTS SIM-o ($0.705$)---and is our default.
Larger $w$ ($\geq 1.5$) marginally trades intelligibility against SIM-o and UTMOS on Seed-TTS.

\section{Streaming Efficiency}
\label{app:streaming_efficiency}
 
This appendix provides the full streaming efficiency numbers summarized in \Cref{sec:results}.
 
\paragraph{Comparability Caveat}
The Qwen3-TTS~\cite{hu2026qwen3} numbers reported below are taken from its own technical report, not measured by us.
The report does not fully specify the GPU, optimization, batching, or concurrency level used, and the systems also differ in model size and token rate (12\,Hz vs.\ 25\,Hz variants).
We therefore do \emph{not} treat the two sets of numbers as a controlled, like-for-like comparison, and we avoid drawing head-to-head speedup conclusions from them.
We include the Qwen3-TTS numbers only as a loose reference point---the best publicly available figures for a comparable streaming AR TTS system---to situate our measurements in context.
All Chatterbox-Flash numbers are measured by us under the protocol below.
 
\paragraph{Measurement Protocol}
Both metrics are computed on $50$ utterances at concurrency $1$, measured up to the moment the first audio packet is emitted by the server.
\emph{Time to first packet} (TTFP) is the wall-clock time from receiving the request to emitting the first audio packet; \emph{real-time factor} (RTF) is the ratio of generation wall-clock time to synthesized audio duration.
All Chatterbox-Flash numbers are measured on NVIDIA H100 GPUs using FlashInfer attention kernels and paged key-value cache management (\Cref{app:impl}).
 
\paragraph{Results}
Under the efficiency-oriented configuration ($D = 16$, $\alpha = 0.5$), Chatterbox-Flash attains TTFP $118$~ms and RTF $0.107$ (\Cref{tab:streaming_efficiency_appendix}).
Its measured TTFP falls in a similar range to the Qwen3-TTS figures ($97$--$150$~ms across the four reported variants), and its RTF is numerically lower than all of them ($0.234$--$0.313$ as reported), corresponding to roughly $9\times$ real-time synthesis at concurrency $1$ in our setup.
More aggressive settings reduce these numbers further: at $D = 32$, $\alpha = 0.75$, Chatterbox-Flash records TTFP $103$~ms and RTF $0.076$ ($\sim$$13\times$ real time).
Because the two systems were measured under different and not fully specified conditions, we report these values side by side to indicate that Chatterbox-Flash operates in a competitive latency--throughput regime, rather than to claim a controlled speedup over Qwen3-TTS.

\begin{table}[!h]
\centering
\small
\setlength{\tabcolsep}{4pt}
\renewcommand{\arraystretch}{1.05}
\begin{tabular}{l c c c}
\toprule
\textbf{Config} & \textbf{SIM-o}$\uparrow$ & \textbf{WER}$\downarrow$ & \textbf{UTMOS}$\uparrow$ \\
\midrule
$D = 16$, $\alpha = 0.5$ (default) & 0.688 & 2.03 & 4.07 \\
$D = 16$, $\alpha = 0.75$          & 0.687 & 2.27 & 4.06 \\
$D = 24$, $\alpha = 0.5$           & 0.687 & 2.63 & 4.07 \\
$D = 24$, $\alpha = 0.75$          & 0.688 & 2.74 & 4.07 \\
$D = 32$, $\alpha = 0.5$           & 0.686 & 3.86 & 4.06 \\
$D = 32$, $\alpha = 0.75$          & 0.685 & 3.70 & 4.05 \\
\midrule
\multicolumn{4}{l}{\textit{Offline reference (\Cref{tab:off_str_quality})}} \\
$D = 16$, $\alpha = 0.5$           & 0.713 & 1.67 & 4.28 \\
\bottomrule
\end{tabular}
\caption{Production streaming-server quality on LibriSpeech-PC test-clean, under the same configurations as \Cref{tab:streaming_efficiency_appendix} (chunk vocoder + early-emit enabled). Unlike the decoder-only comparison in \Cref{tab:off_str_quality}, these numbers include server-level sampling optimizations and vary block size, so a modest quality drop versus offline appears, growing in WER at larger $D$.}
\label{tab:streaming_quality}
\end{table}

\subsection{Streaming Server Quality}
\label{app:streaming_quality}

\Cref{tab:streaming_quality} reports objective quality measured on the production streaming server used for the latency numbers above (\Cref{tab:streaming_efficiency_appendix}), i.e.\ with the chunk-wise vocoder and early-emit block serving enabled.
This differs from the controlled comparison of \Cref{tab:off_str_quality}, which isolates the decoder at a fixed configuration: the production server additionally varies block size and applies sampling-level optimizations, so its numbers reflect end-to-end deployment behavior and can differ from the decoder-only measurements.
At the default $D = 16$, $\alpha = 0.5$, the server attains SIM-o $0.688$, WER $2.03$, and UTMOS $4.07$ on LibriSpeech-PC test-clean.
Across configurations, SIM-o and UTMOS stay stable while WER grows with block size, consistent with the block-size ablation (\Cref{fig:block_steps}); this is the main quality cost of the most aggressive latency settings.

\paragraph{Vocoder Chunk Schedule}
The chunk-wise vocoder operates with a progressively widening chunk schedule.
The first chunk has a duration of $0.46$~s ($\sim$$12$ tokens at the $25$~Hz codec rate), with each subsequent chunk grown by a factor of $5.0$, capped at $6.0$~s ($150$ tokens).
A small initial chunk reduces TTFP, since the vocoder can emit the first audio packet as soon as a short window of speech tokens is available; subsequent larger chunks amortize vocoder overhead during sustained synthesis.
The chunk duration can be tuned per deployment scenario, with shorter initial chunks reducing TTFP at the cost of slightly higher steady-state cost.

\paragraph{Early-Emit Block Serving}
We further reduce streaming latency through an \emph{early-emit} optimization at the block boundary.
During block decoding, masked positions within a block are unmasked progressively, and the early-decoding schedule (see \nameref{par:inference_early} paragraph) typically commits positions in a non-contiguous order.
Whenever the committed positions form a contiguous left-aligned prefix of the block---that is, the remaining masked positions occupy only the right portion---we emit the prefix tokens to the vocoder immediately, without waiting for the entire block to finish decoding.
This overlaps vocoder decoding of the early-committed tokens with the remaining denoising steps for the same block, yielding additional latency reduction while preserving the block-by-block streaming abstraction.

\section{Implementation Details}
\label{app:impl}

The inference engine is built on FlashInfer, with three customizations specific to block-diffusion TTS that differ from standard LLM serving.

\paragraph{Hybrid Causal/Non-Causal Attention}
Standard LLM serving uses causal attention throughout.
Our inference loop instead interleaves two attention patterns within a single decoding step: the conditioning prefix is encoded with causal attention and cached once, while the current block is decoded with non-causal attention so that all masked positions in the block attend bidirectionally to each other (\Cref{sec:inference_blockdec}).
The two patterns share the same paged key-value buffer and differ only in the causal flag passed to the attention kernel.

\paragraph{Frozen Prefix, Growing Block Cache}
The prefix forward is invoked once per generation and writes the prefix key-value entries into the paged buffer.
Subsequent block forwards reuse this frozen prefix cache, append the current block's key-value entries, and never recompute prefix attention.
This realizes the two-cache abstraction described in \Cref{sec:inference_blockdec}.

\paragraph{Cache Snapshot for CFG}
Classifier-free guidance (\Cref{sec:inference_cfg}) requires both a conditional and an unconditional forward at each block step.
We snapshot the paged cache before the unconditional pass and restore it afterwards, so the conditional state is preserved without recomputing the prefix.

\paragraph{CUDA Graph Replay}
Because the block size $D$ is fixed at inference time, the per-step forward---attention, feed-forward, and the speech head---is captured as a single CUDA graph and replayed for every step of every block.
This removes the per-launch overhead that otherwise dominates short-sequence, small-batch inference.

\section{Block-Size Scaling Explorations}
\label{app:block_scaling}

This appendix summarizes alternative block-diffusion configurations we explored when attempting to scale beyond the block sizes used in our main results.
None of these alternatives delivered both stable training at large block sizes and the prosodic quality required for production deployment, but they characterize where the difficulty lies and motivate future work.

\paragraph{Fully Causal Block Formulation (CARD-style)}
Following the causal autoregressive diffusion formulation of CARD~\cite{ruan2026causalautoregressivediffusion}, we trained a variant in which all attention within and across blocks is strictly causal, removing the intra-block bidirectional context of our default formulation.
At small block sizes ($D \leq 4$), this variant produces intelligible speech and offers the fastest inference among the configurations we tested, since a single forward pass per block is sufficient and no iterative refinement is required.
Starting from $D = 5$, however, generated speech begins to exhibit prosodic collapse and over-smoothing, with reduced pitch and energy variation and noticeably flatter intonation.
The fully causal training signal appears to be too weak to support meaningful parallelism inside the block, suggesting that the bidirectional intra-block context used by our main formulation is necessary for prosodic fidelity in TTS.

\paragraph{Block-Size Annealing with Self-Distillation}
Inspired by A2D-VL~\cite{arriola2025ar2d}, which gradually increases the diffusion prediction window during fine-tuning, we experimented with a data-free self-distillation recipe that anneals the block size starting from $D = 1$ (purely autoregressive) and progressively widening the block.
This approach was able to maintain coherent speech up to roughly $D = 8$.
Beyond this point, the model's per-position confidence collapses to a narrow range, and sampling becomes highly sensitive to temperature and CFG scale---small perturbations in the sampling configuration produce qualitatively different outputs.
Extending the schedule to larger block sizes did not produce a stable regime within our compute budget.

\paragraph{Discussion}
Across these explorations, two distinct failure modes recur: (i) loss of prosodic variability when the training signal does not enforce bidirectional intra-block context, and (ii) collapsed confidence and high sampling sensitivity when block size grows faster than the model's denoising capability adapts.
Our default block-diffusion formulation with bidirectional intra-block attention and a moderate block size offered the most stable trade-off among the recipes we tested, but a principled approach to scaling block size while preserving both prosody and confidence calibration remains an open question.

\section{Streaming Capability of Baselines}
\label{app:streaming_baselines}

Our main results compare Chatterbox-Flash against NAR baselines that, to our knowledge, are not designed for streaming inference.
In particular, OmniVoice~\cite{zhu2026omnivoice} is formulated as a full-sequence masked diffusion model whose architecture does not natively support streaming.
The authors have indicated\footnote{\url{https://github.com/k2-fsa/OmniVoice/issues/6}} that a form of \emph{pseudo-streaming} can nonetheless be obtained by splitting the input text into smaller chunks and processing them sequentially.

We did not evaluate this configuration in the present work for two reasons.
First, chunked sampling deviates from the model's training-time assumption of full-sequence bidirectional context, and the resulting quality--latency trade-off would likely depend on chunking strategy, overlap, and re-tuning of the denoising schedule.
Second, any number reported under such an adaptation would reflect our specific reimplementation rather than the original system, making a fair comparison difficult.

We view a controlled study of pseudo-streaming OmniVoice---and, more broadly, of how full-sequence diffusion models can be retrofitted for streaming---as a meaningful direction for future work, as it would help disentangle whether the streaming advantage observed in our setting stems from the block-diffusion formulation itself or from training-time alignment with the streaming inference pattern.

\end{document}